\documentclass[12pt]{article}
\usepackage{mathrsfs}
\usepackage{natbib,graphicx,subfigure,setspace,lscape,longtable}
\usepackage{natbib,epsfig,graphicx}
\usepackage{mathrsfs,amsmath,amsthm,amssymb,color}
\usepackage{algorithm}
\usepackage{algorithmic}

\bibpunct{(}{)}{;}{a}{,}{,}

\newtheorem{theorem}{Theorem}
\newtheorem{lemma}{Lemma}
\newtheorem{proposition}{Proposition}
\newcommand{\csection}[1]
    {\begin{center}
        \stepcounter{section}
        {\bf\large\arabic{section}. #1}
    \end{center}
    \vspace{-0.15 cm}
}

\newcommand{\scsection}[1]
    {\begin{center}
        {\bf\large #1}
    \end{center}
    \vspace{-0.15 cm}
}

\newcommand{\csubsection}[1]{
\vspace{-0.25 cm}
\begin{center}
\stepcounter{subsection}
{\it\arabic{section}.\arabic{subsection}. #1}
\end{center}
\vspace{-0.25 cm}
}

\newcommand{\scsubsection}[1]{
\vspace{-0.25 cm}
\begin{center}
\stepcounter{subsection}
{\it #1}
\end{center}
\vspace{-0.25 cm}
}

\def\beq{\begin{equation}}
\def\eeq{\end{equation}}
\def\beqr{\begin{eqnarray}}
\def\eeqr{\end{eqnarray}}
\def\beqrs{\begin{eqnarray*}}
\def\eeqrs{\end{eqnarray*}}
\def\bet{\begin{theorem}}
\def\eet{\end{theorem}}
\def\bel{\begin{lemma}}
\def\eel{\end{lemma}}
\def\bep{\begin{proposition}}
\def\eep{\end{proposition}}
\def\bg{\begin{figure}[tbph]\begin{center}}
\def\eg{\end{center}\end{figure}}

\def\bc{\begin{center}}
\def\ec{\end{center}}

\def\wh{\widehat}

\def\var{\mbox{var}}
\def\cov{\mbox{cov}}

\def\mR{\mathbb{R}}
\def\mB{\mathcal{B}}
\def\mS{\mathbb{S}}
\def\mX{\mathbb{X}}

\numberwithin{equation}{section}

\begin{document}

\begin{center}

{\bf\Large A Sequential Addressing Subsampling Method for Massive Data Analysis under Memory Constraint}\\

Rui Pan$^{1}$, Yingqiu Zhu$^{2}$, Baishan Guo$^{3}$, Xuening Zhu$^{4}$, and Hansheng Wang$^{5}$\\
{\it $^1$ Central University of Finance and Economics, China. \\
$^2$ Renmin University of China, China. $^3$ Facebook Inc., USA. \\
$^4$ Fudan University, China. $^5$ Peking University, China.}

\end{center}

\begin{footnotetext}[1]
{ 	
The research of Rui Pan is supported in part by National Natural Science Foundation of China (NSFC. 11601539, 11631003, 71771224), Program for Innovation Research in Central University of Finance and Economics. The research of Xuening Zhu is supported by the National Natural Science Foundation of China (NSFC, 11901105, 71991472, U1811461), the Shanghai Sailing Program for Youth Science and Technology Excellence (19YF1402700), and the Fudan-Xinzailing Joint Research Centre for
Big Data, School of Data Science, Fudan University.
The research of Hansheng Wang is supported in part by National Natural Science Foundation of China (NSFC. 11831008, 11525101, 71532001). }
\end{footnotetext}
\begin{footnotetext}[2]
{Yingqiu Zhu is the corresponding author.}
\end{footnotetext}

\begin{singlespace}
\begin{abstract}

The emergence of massive data in recent years brings challenges to automatic statistical inference. This is particularly true if the data are too numerous to be read into memory as a whole. Accordingly, new sampling techniques are needed to sample data from a hard drive. In this paper, we propose a sequential addressing subsampling (SAS) method, that can sample data directly from the hard drive. The newly proposed SAS method is time saving in terms of addressing cost compared to that of the random addressing subsampling (RAS) method. Estimators (e.g., the sample mean) based on the SAS subsamples are constructed, and their properties are studied. We conduct a series of simulation studies to verify the finite sample performance of the proposed SAS estimators. The time cost is also compared between the SAS and RAS methods. An analysis of the airline data is presented for illustration purpose.\\


\end{abstract}
\end{singlespace}

\newpage

\csection{INTRODUCTION}

In the past decade, massive data are becoming increasingly prevalent in many fields including finance, economics, marketing, and many others.
This brings challenges to automatic statistical inference. In this case, the popularly used bootstrap method \citep{Efron:1979} is computationally too expensive or even infeasible.
Accordingly, various subsampling methods have been proposed. The key difference is that the data size resampled by subsampling methods is typically much smaller than that of traditional bootstrap methods, which makes computation practically feasible for massive data. Important studies in this regard include but are not limited to the $m$ out of $n$ bootstrap \citep{Bickel:1997}, the Bag of Little Bootstrap (BLB, Kleiner et~al., 2014), the subsampling double bootstrap (Sengupta et~al., 2016) and many others.

In practice, if the dataset is small and thus can be read into the computer memory as a whole, the computational cost is mainly associated with in-memory processing and has little to do with the hard drive.
However, if the data are too large to be read into memory as a whole, they have to be stored on the hard drive. To understand this, consider the {\it airline} data ({\it http://stat-computing.org/dataexpo/2009/}) used in this study as an example. The dataset contains more than one hundred million air flight records. The fully scaled data with all dummy variables having been generated are more than 200G in size. Accordingly, the data can be easily stored on the hard drive but are very difficult to read into memory as a whole. Consequently, if subsampling methods are to be applied to this dataset, the computer needs to repeatedly resample data from the hard drive instead of in the computer memory.

It is notable that the time needed to sample one data point from the hard drive is considerably longer than that required in the case of memory. More specifically, the time cost mainly consists of two parts. The first is related to the time needed to identify the target data point on the hard drive. We refer to this part as the {\it addressing cost}. The second is related to the time needed to read the target data point into memory. We refer to this part as the {\it I/O cost}. To summarize, we refer to the addressing cost as well as the I/O cost together as the {\it hard drive sampling cost} (HDSC), which is the time needed to retrieve one particular data point from the hard disk into the computer memory. Intuitively, HDSC could be negligible if the size of the raw data is relatively small. However, our experiments suggest that HDSC could be significant if the raw data are massive. To our best knowledge, there has been little discussion of this aspect in existing studies.

To develop our idea, we conduct a simple simulation on our workstation, which is a Windows 10 computer with a 3.4 GHz Intel Core i7-6700 processor and 16GB random access memory. We simulate a raw dataset with 200 million observations independently from a standard normal distribution. For each data point, three decimal places are kept, which leads to a 1.39G raw dataset. The data are then stored on the hard drive. In each simulation replication, we randomly select one observation from the raw data and then read it into memory. We repeat this process one million times. Obviously, this is a standard sampling method and we refer to it as the {\it random addressing subsampling} (RAS) method. Thereafter, the HDSC associated with the RAS method is recorded. Once the selected subsample has been read into memory, the sample mean is computed, and the corresponding computation time is also recorded. The experiment is randomly replicated 100 times. The average HDSC is 45.85 seconds, while the average computation time is only 0.0083 seconds. The former is approximately 6 thousand times as long as the latter, which suggests that HDSC represents a much greater time cost than the computation time. Consequently, HDSC cannot be ignored when we apply subsampling methods to massive datasets. However, most existing subsampling methods are quite inefficient in terms of HDSC, and little research in this regard exists. Thus, developing a novel subsampling method that is efficient in terms of HDSC becomes a problem of great importance.

To solve this problem, we develop in this paper a {\it sequential addressing subsampling} (SAS) method. The newly developed sampling method is very efficient in terms of the HDSC and consists of two steps, namely, a preparation step and a sequential sampling step. In the preparation step, a random shuffling operation is performed on the raw data. The aim of shuffling is to randomly sort the raw data and keep them on the hard drive for the following sequential sampling step. The random shuffling operation we propose is not time consuming. For instance, a shuffle of the raw airline data (approximately 10G) takes less than 200 seconds. Thus, it is practically acceptable. Once the preparation step has been completed, the sequential sampling step can be performed repeatedly based on the shuffled data to obtain subsamples. This step contains two substeps. In the first substep, one data point should be randomly identified on the hard drive as a starting point. This operation incurs a considerable addressing cost but is only performed for one data point. Once the starting point has been fixed, data are sampled sequentially in the second substep. Because the remaining data points are sampled sequentially, no additional addressing cost is required, and the associated I/O cost is considerably reduced. This situation substantially reduces the sampling cost. To assess this idea, we repeat the former experiment but replace the RAS method by SAS. The average HDSC decreases significantly from 45.85 seconds to 1.11 seconds.

Based on the subsamples generated by SAS, various standard statistics can be computed. Such statistics include but are not limited to the sample moments and their smooth transformations. For convenience, we refer to one particular estimate computed based on one subsample as a subsample estimate. Since multiple subsamples are to be generated by SAS, we should obtain multiple subsample estimates. They are then averaged so that a combined and more accurate estimate can be produced. It is then referred to as the SAS estimator. Due to the sequential sampling nature of SAS, the asymptotic properties of the resulting SAS estimator are different from those associated with the RAS method. In this regard, we develop a relatively complete asymptotic theory for the proposed SAS method. The inference procedure based on the SAS method is also developed.

To summarize, the contributions of our study entail four aspects. First, we focus on HDSC, which has been rarely discussed in the past literatures especially under memory constraints. We emphasize this important issue and call for more studies. Second, we develop a brand-new sequential subsampling method that can sample data directly from the hard drive. It is proved to be time saving in terms of the addressing cost and excellent in terms of computational feasibility. As a result, the newly proposed SAS method is a powerful tool if data cannot be read into memory as a whole but are stored on the hard drive. Third, we develop new statistics based on SAS subsamples. The statistics include sample moments and their smooth transformations. The corresponding statistical properties are also obtained. Fourth, the nature of SAS also allows for automatic inference. It is particularly useful if the standard error is too complicated to derive (e.g., the standard error of the sample correlation).

The rest of the article is organized as follows. In Section 2, the SAS method is proposed. For a complete comparison, we also illustrate the algorithm of the RAS method. In Section 3, we present the theoretical results of SAS estimators. Section 4 describes a series of simulation studies conducted to verify the finite sample performance of the proposed SAS estimators. The HDSC values are also compared for the SAS and RAS methods. A real-world example of airline data is presented in Section 5. All technical proofs are left in the appendix.

\csection{SEQUENTIAL ADDRESSING SUBSAMPLING}

\csubsection{The Random Addressing Subsampling}

As discussed above, a dataset of interest is often too large to be read into memory as a whole. However, it can be stored on the hard drive. As a result, data need to be resampled from the hard drive instead of memory. In this subsection, we present the RAS method. More specifically, random addressing refers to moving the pointer to an arbitrary address on the hard drive and accessing the data point at that address. Given a dataset consisting of $N$ data points, to retrieve $n$ data points as a subsample, the RAS method needs to conduct random addressing $n$ times and concatenate $n$ data points obtained from the file. Since only the accessed data will be loaded into memory, the required storage of memory is reduced from $O(N)$ to $O(n)$. While conducting subsampling, we keep the raw data on the hard drive instead of loading the whole dataset into memory. The RAS method is further illustrated in detail in Algorithm \ref{al:random}.

\begin{algorithm}[ht]
\caption{Random addressing subsampling}
\begin{algorithmic}
\STATE \textbf{Input}: $n$: subsample size; $F$: file containing the original dataset; $n_f$: size of $F$, measured in bytes.
\STATE $count \leftarrow 0$;
\STATE $S \leftarrow \{\}$;
\REPEAT
\STATE Randomly select an integer $p_b$ from $\{0,1,\cdots, n_f\}$;
\STATE Execute addressing on $F$ using $p_b$ as the offset. More specifically, move the pointer to $p_b$ from the header of $F$;
\IF{there exists the next line}
\STATE Move the pointer to the header of the next line;
\ELSE
\STATE Move the pointer to the header of $F$;
\ENDIF
\STATE Read line $l$;
\STATE $S \leftarrow S \bigcup \{l\}$;
\STATE $count \leftarrow count + 1$;
\UNTIL $count = n$.
\STATE \textbf{Output}: subample $S$ consisting of $n$ lines.
\end{algorithmic}\label{al:random}
\end{algorithm}

\noindent{\bf Remark 1.} Once the offset has been specified, the random addressing can be performed in a very short time. However, the offset has to be defined in terms of bytes instead of the number of lines. It is unlikely to match exactly an address of the header of a line. Therefore, after performing the random addressing, we have to move the pointer to the header of the next line. This operation is implemented via detection of line breaks. Then, the algorithm is ready to read a non-truncated line.

\csubsection{The Sequential Addressing Subsampling}

As shown in the Introduction, selecting a sample randomly from the hard drive into the computer memory is time consuming. To reduce the time cost incurred by the RAS method, we develop a brand new addressing based subsampling method called the SAS method. More specifically, to retrieve $n$ data points as a subsample, the SAS method requires random addressing to be performed only once. Nevertheless, this approach requires a shuffling operation described in the following subsection to be executed on the raw dataset as preparation. The shuffling operation leads to a randomly rearranged dataset. Using the shuffled dataset, the SAS method can be executed repeatedly. The details of SAS are shown in Algorithm \ref{al:sequence}.

\begin{algorithm}[h]
\caption{Sequential addressing subsampling}
\begin{algorithmic}
\STATE \textbf{Input}: $n$: subsample size; $F_s$: file containing the shuffled dataset; $n_f$: file size.
\STATE $count \leftarrow 0$;
\STATE $S \leftarrow \{\}$;
\STATE Randomly select an integer $p_b$ from $\{0,1,\cdots, n_f\}$;
\STATE Execute addressing on $F_s$ using $p_b$ as the offset. More specifically, move the pointer to $p_b$ from the header of $F_s$;
\REPEAT
\IF{there exists the next line}
\STATE Move the pointer to the header of the next line;
\ELSE
\STATE Move the pointer to the header of $F_s$;
\ENDIF
\STATE Read line $l$;
\STATE $S \leftarrow S \bigcup \{l\}$;
\STATE $count \leftarrow count + 1$;
\UNTIL $count = n$.
\STATE \textbf{Output}: sample $S$ consisting of $n$ lines.
\end{algorithmic}\label{al:sequence}
\end{algorithm}

\noindent{\bf Remark 2.} The SAS method requires only a sigle execution of random addressing. Thus, the time cost is supposed to be much lower than that of the RAS method. As mentioned in the Introduction, the time cost of shuffling can be well controlled. As a result, for cases in which the subsampling has to be conducted repeatedly, the SAS method is considered to be much more efficient than the RAS method.

\csubsection{Shuffle}

The ASA method requires a shuffling operation. We perform it as follows.

\noindent \textbf{Step 1}: Initialize $b$ empty cache files. For each line of $F$, randomly assign its index (e.g., the byte position of that line) to one of $b$ cache files. This procedure could be implemented using sequential reading of $F$. During the reading process, only the read line is loaded into memory, and it is released immediately after the random assignment. Thus, the memory required by this step is $O(1)$.

\noindent \textbf{Step 2}: Generate a random permutation of sequence $\{1, \cdots, b\}$. Based on the permutation, $b$ cache files are ordered as $\{f_{(1)}, \cdots, f_{(b)}\}$.

\noindent \textbf{Step 3}: Initialize an empty file $F^*$ to store the shuffled data. For each cache file $f_{(i)}$ with $1\leq i\leq b)$, we conduct the following two operations.
\begin{itemize}
\item First, load the cache file into memory and shuffle the content. Since the content of each cache file consists of indexes, and occupies much less space than the original data, one can greatly reduce the required memory.
\item Second, for the shuffled cache file, sequentially read each index and then obtain the corresponding line in $F$ according to the index. The retrieval of the corresponding line in $F$ is implemented using addressing on the hard drive. Only the retrieved line is input into memory. Therefore, the required memory is $O(1)$.
Once a line in $F$ has been read, it is added to $F^*$.
\end{itemize}
Once all cache files have been processed, we can obtain a completely shuffled dataset $F^*$.

\csection{THEORETICAL RESULTS}

\csubsection{Notations}

Let $\mX=\{X_1,\cdots,X_N\}$ be the collection of all samples, where $N$ is sample size. Assume that $X_1,\cdots,X_N$ are the samples after shuffling. Furthermore, assume that $E(X_i)=\mu$, $\var(X_i)=\sigma^2$, and $E(X_i-\mu)^4=\gamma\sigma^4$ for $\gamma\geq 1$. Define $\bar{X}=N^{-1}\sum_{i=1}^N X_i$, where $E(\bar{X})=\mu$ and $\mbox{SE}(\bar{X})=\sigma/\sqrt{N}$.
 Using the procedure of the SAS method, $N-n+1$ sequential subsamples can be generated, where $n$ is the subsample size. For convenience, let $K=N-n+1$. More specifically, denote the $k$th sequential subsample by $\mX_{(k)}=\{X_k,X_{k+1},\cdots,X_{k+n-1}\}$, and the associated sample mean can be defined as $\bar X_{k}=n^{-1}\sum_{i=k}^{k+n-1} X_i$. We further denote all sample means of sequential subsamples by $\mB=\{\bar X_{k}:1\leq k\leq K \}$. Furthermore, define $\bar{\bar{X}}=K^{-1}\sum_{k=1}^K \bar X_{k}$. If one can obtain all sequential subsamples, then $\bar{\bar{X}}$ can be calculated. We study the properties of $\bar{\bar{X}}$ for theoretical completeness even though it is unrealistic to obtain all sequential subsamples. The properties of $\bar{\bar X}$ are presented in the next lemma.

\bel
Define $\bar{\bar{X}}=K^{-1}\sum_{k=1}^K \bar X_{k}$, where $K=N-n+1$. Further, assume that $n\rightarrow\infty$, $N\rightarrow\infty$, and $n/N\rightarrow 0$. Then, the expectation and variance of $\bar{\bar{X}}$ are
$E(\bar{\bar{X}})=\mu$ and
\beqr\label{eq:barbarx}
\var\big(\bar{\bar X}\big)&=&\frac{N-n+1+(n-1)(3N-4n+2)/3}{n(N-n+1)^2}\sigma^2\\
&=&\frac{\sigma^2}{N}\Bigg\{1+\frac{2n}{3N}+o\Big(\frac{n}{N}\Big)\Bigg\}.\nonumber
\eeqr
\eel
\noindent
The proof of Lemma 1 is provided in Appendix A. Lemma 1 shows that $\bar{\bar X}$ is an unbiased estimator of the population mean $\mu$. Furthermore, the leading order of its variance is $O(N^{-1})$, where $N$ is the total number of observations. The higher order of $\var(\bar{\bar X})$ is $O(nN^{-2})$. This term declines if subsample size $n$ decreases or the total sample size $N$ increases. More specifically, if $n$ is too large, there will be more overlaps among the sequential subsamples, leading to the increase in $\var(\bar{\bar X})$.

\csubsection{The Sample Mean}

In practice, one can randomly obtain $B$ sequential subsamples by the SAS method. Furthermore, denote the sample means resulting from $B$ sequential subsamples by $\mS=\{\bar{X}_{(1)},\cdots,\bar{X}_{(b)},\cdots,\bar{X}_{(B)}\}$. Define the sample mean derived from the SAS method as
\beq\label{eq:barXB}
\bar{\bar{X}}_B=B^{-1}\sum_{b=1}^B \bar{X}_{(b)},
\eeq
which is the statistic of interest. We next study the expectation and variance of $\bar{\bar{X}}_B$.
First,
$E(\bar{\bar{X}}_B)=E(\bar{X}_{(b)})=\mu$. As a result, $\bar{\bar{X}}_B$ is an unbiased estimator of $\mu$.
As to the variance of $\bar{\bar{X}}_B$, we state the following theorem.
\bet
Assume $B\rightarrow\infty$, $n\rightarrow\infty$, $N\rightarrow\infty$, and $n/N\rightarrow 0$. Then, we have that
\beq
\var(\bar{\bar X}_B)=\sigma^2\bigg(\frac{1}{nB}+\frac{1}{N}\bigg)\big\{1+o(1)\big\}.
\eeq
\eet
\noindent
The proof of Theorem 1 is provided in Appendix B. Theorem 1 shows that the variance of $\bar{\bar X}_B$ is determined by two main terms. The first arises from the variance of $\bar{\bar X}$, where the leading term is $\sigma^2/N$ and $N$ is the total sample size. This term is due to the fundamental random error generated by the whole sample. Consequently, it cannot be eliminated by increasing the number of subsamples $B$ or subsample size $n$. The second term arises from the sampling procedure, which is $\sigma^2/(nB)$. As long as we increase the subsample size or obtain more subsamples, the term $nB$ can be larger than $N$. In this scenario, the leading order of $\var(\bar{\bar X}_B)$ becomes $O(N^{-1})$, and the increase in $nB$ will not lead to a further decrease in the variance of $\bar{\bar X}_B$. From the subsampling point of view, a large $nB$ will increase the sampling cost. As a result, in real practice, $nB$ should be smaller than $N$. We will try various combinations of $n$ and $B$  in the simulation studies.

\csubsection{The Inference of Sample Mean}

To perform automatic inference, i.e., to estimate $\var(\bar{\bar X}_B)$, we construct the following statistic based on the SAS procedure,
\beq
\wh{\mbox{SE}}^2(\bar{\bar X}_B) = \frac{n}{B-1}\Big(\frac{1}{nB}+\frac{1}{N}\Big) \sum_{b=1}^B \Big(\bar{X}_{(b)}-\bar{\bar{X}}_{B}\Big)^2
=\frac{c}{B-1}\sum_{b=1}^B \Big(\bar{X}_{(b)}-\bar{\bar{X}}_{B}\Big)^2,\label{eq:SE}
\eeq
where $c=n\{(nB)^{-1}+N^{-1}\}$ can be regarded as a scaler. The properties of $\wh{\mbox{SE}}^2$ are presented in the following theorem.
\bet
Assume that $n\rightarrow\infty$, $N\rightarrow\infty$, $B\rightarrow\infty$, and $n/N\rightarrow 0$. Then we have that
\beqrs
E\Big\{\wh{\mbox{SE}}^2(\bar{\bar X}_B)\Big\}&=&\sigma^2\Big(\frac{1}{nB}+\frac{1}{N}\Big)\big\{1+o(1)\big\}.\\
\var(\bar{\bar X}_B)-E\Big\{\wh{\mbox{SE}}^2(\bar{\bar X}_B)\Big\}&=&O\Big(\frac{n}{N^2}\Big).
\eeqrs
\eet
\noindent
The proof of Theorem 2 is presented in Appendix C. It can be seen that the leading order of $\var(\bar{\bar X}_B)$ and the expectation of $\wh{\mbox{SE}}^2(\bar{\bar X}_B)$ are the same. However, as an estimator of $\var(\bar{\bar X}_B)$, $\wh{\mbox{SE}}^2(\bar{\bar X}_B)$ is biased. The order of the bias is $O(nN^{-2})$, which decreases as $N$ increases or as $n$ decreases.
In the following theorem, we analyze the variance of $\wh{\mbox{SE}}^2(\bar{\bar X}_B)$.
\bet
Assume that $B\rightarrow\infty$, $n\rightarrow\infty$, $N\rightarrow\infty$, and $n/N\rightarrow 0$. Then we have that
\beqr
\var\Big\{\wh{\mbox{SE}}^2(\bar{\bar X}_B)\Big\}=2\sigma^4\Big(\frac{1}{nB}+\frac{1}{N}\Big)^2
\Bigg\{O\Big(\frac{1}{B}\Big)+O\Big(\frac{1}{N}\Big)+O\Big(\frac{n^2}{N^2}\Big)\Bigg\}.
\eeqr
\eet
\noindent
The proof of Theorem 3 is provided in Appendix D. To make the variance of $\wh{\mbox{SE}}^2(\bar{\bar X}_B)$ as small as possible, one is expected to make $B$, the number of selected sequential subsamples, as large as possible.

\csubsection{General Statistics}

In the preceding subsections, we estimate the population mean $\mu$ and make the inference. Additionally, we are often interested in a transformation of $\mu$, denoted by $\theta=g(\mu)$ where $g(\cdot)$ is a continuously differentiable function. To estimate $\theta$, we develop the following two estimators and study their properties.

First, we develop a straightforward estimator. Since $\bar{\bar X}_B$ is used for estimating $\mu$, it is natural to use $g(\bar{\bar X}_B)$ to estimate $\theta$ and we denote it by $\tilde{\theta}_B$.
Using a standard Taylor expansion, we obtain
$$\tilde\theta_B=g(\bar{\bar X}_B)=g(\mu)+\dot{g}(\mu)(\bar{\bar X}_B-\mu)+\ddot{g}(\xi_B)(\bar{\bar X}_B-\mu)^2,$$
where $\dot{g}(\cdot)$ and $\ddot{g}(\cdot)$ are the first- and second-order derivatives of $g(\cdot)$,
and $\xi_B$ is between $\bar{\bar X}_B$ and $\mu$.
Second, for the $k$th sequential subsample, $g(\mu)$ can be estimated by $g(\bar X_k)$. Recall that one can obtain $K=N-n+1$ sequential subsamples, and in practice, we often select $B$ sequential subsamples to form an estimator. Hence, another possible estimator of $\theta$ is \beq\label{eq:hatB}
\hat\theta_B=B^{-1}\sum_{b=1}^B g\big(\bar X_{(b)}\big).
\eeq
Similarly, we have the following Taylor expansion,
\beq
g(\bar{X}_k)=g(\mu)+\dot{g}(\mu)(\bar X_k-\mu)+\ddot{g}(\xi_k)(\bar X_k-\mu)^2,\label{eq:TEbarxk}
\eeq
where $\xi_k$ is between $\bar X_k$ and $\mu$. As a result, from (\ref{eq:hatB}) and (\ref{eq:TEbarxk}) we obtain
$$\hat\theta_B=g(\mu)+\dot g(\mu)B^{-1}\sum_{b=1}^B(\bar X_{(b)}-\mu)+B^{-1}\sum_{b=1}^B\ddot{g}(\xi_{(b)})(\bar X_{(b)}-\mu)^2,$$
where $\xi_{(b)}$ is between $\bar X_{(b)}$ and $\mu$.

To study the properties of $\tilde\theta_B$ and $\hat\theta_B$, we further state the following conditions.
\begin{itemize}
\item [(C1)] Assume that $g(\cdot)$ is a continuously differentiable function with $|\ddot g(\cdot)|\leq C_{\max}$,
where $C_{\max}$ is some positive constant.
\item [(C2)] Assume that $\sqrt{N}=o(n)$.
\end{itemize}
We next present the properties of $\tilde\theta_B$ and $\hat\theta_B$ in the next theorem.

\bet
Assuming conditions (C1) and (C2), we then have
\beqrs
E(\tilde\theta_B)&=&g(\mu)+\sigma^2O\Big(\frac{1}{nB}+\frac{1}{N}\Big),~\var(\tilde\theta_B)=\dot g^2(\mu)\sigma^2\Big(\frac{1}{nB}+\frac{1}{N}\Big)\big\{1+o(1)\big\}.\\
\\
E(\hat\theta_B)&=&g(\mu)+\sigma^2O\Big(\frac{1}{n}\Big),~\var(\hat\theta_B)=\dot g^2(\mu)\sigma^2\Big(\frac{1}{nB}+\frac{1}{N}\Big)\big\{1+o(1)\big\}.
\eeqrs
\eet
\noindent
The proof of Theorem 4 is provided in Appendix E. Theorem 4 shows that both $\tilde\theta_B$ and $\hat\theta_B$ are biased.
The bias of $\tilde\theta_B$ is $O(n^{-1}B^{-1}+N^{-1})$, while that of $\hat\theta_B$ is $O(n^{-1})$. The variances of both estimators
are of the same order, which is $O(n^{-1}B^{-1}+N^{-1})$. Although the bias of $\tilde\theta_B$ is smaller than that of $\hat\theta_B$, $\tilde\theta_B$ cannot be used for automatic inference.

We then construct the following statistic for automatic inference of $\hat\theta_B$ based on the SAS procedure,
\beq
\wh{\mbox{SE}}^2(\hat\theta_B)=\frac{n}{B-1}\Big(\frac{1}{nB}+\frac{1}{N}\Big)\sum_{b=1}^B\Big\{g\big(\bar X_{(b)}\big)-\hat\theta_B\Big\}^2
=\frac{c}{B-1}\sum_{b=1}^B \Big\{g\big(\bar X_{(b)}\big)-\hat\theta_B\Big\}^2,\label{eq:SEofg}
\eeq
where $c$ is defined in (\ref{eq:SE}).
The expectation and variance of $\wh{\mbox{SE}}^2(\hat\theta_B)$ are given in the next theorem.
\bet
Assuming conditions (C1) and (C2), we have that
\beqrs
E\Big\{\wh{\mbox{SE}}^2(\hat\theta_B)\Big\}&=&\dot g^2(\mu)\sigma^2\Big(\frac{1}{nB}+\frac{1}{N}\Big)\big\{1+o(1)\big\},\\
\var\Big\{\wh{\mbox{SE}}^2(\hat\theta_B)\Big\}&=&2\dot g^4(\mu)\sigma^4\Big(\frac{1}{nB}+\frac{1}{N}\Big)^2
\Bigg\{O\Big(\frac{1}{B}\Big)+O\Big(\frac{1}{N}\Big)+O\Big(\frac{n^2}{N^2}\Big)\Bigg\}.
\eeqrs
\eet
The proof of Theorem 5 is provided in Appendix F.

\csection{NUMERICAL STUDIES}

\csubsection{Model Setups}

We present five examples in numerical studies. The first example is used to evaluate the performance of the SAS sample mean. The second example is considered to study a function of the population mean. The third to fifth examples entail more complicated quantities, i.e., the coefficient of variation (CV), the linear correlation coefficient, and the linear regression coefficient.

{\sc Example 1.} We first generate $X_1,\cdots,X_N$ independently from the standard normal distribution with $\mu=0$ and $\sigma^2=1$.
 As long as we obtain $B$ sequential subsamples with subsample size $n$, the sample mean $\bar{\bar X}_B$ can be used to estimate $\mu$, and the squared standard error $\wh{\mbox{SE}}^2(\bar{\bar X}_B)$ can be calculated according to (\ref{eq:SE}).

{\sc Example 2.} We generate $X_1,\cdots,X_N$ independently from the normal distribution with $\mu=1$ and $\sigma^2=1$. The parameter of interest is $\theta=g(\mu)=\sin(\mu)$. As a result, the estimator can be calculated as $\hat\theta_B=B^{-1}\sum_{b=1}^B\sin(\bar X_{(b)})$. The squared standard error $\wh{\mbox{SE}}^2(\hat\theta_B)$ is derived according to (\ref{eq:SEofg}). To derive the true variance of $\hat\theta_B$, we calculate $\dot g^2(\mu)=\cos^2(\mu)$ and $\mu=1$.

{\sc Example 3.} In this example, we consider a more complicated parameter, i.e., the CV. More specifically, we generate $X_1,\cdots,X_N$ from a normal distribution with $\mu=1$ and $\sigma^2=1$. The population coefficient of variation is $\theta=\sigma/\mu$. Based on sequential subsamples, the SAS estimator is calculated as $\hat\theta_B=B^{-1}\sum_{b=1}^B S_{(b)}/\bar X_{(b)}$, where $S_{(b)}$ is the sample standard deviation based on the $b$th sequential subsample. The squared standard error of $\hat\theta_B$ is calculated according to (\ref{eq:SEofg}). Note that in this example, the true variance of $\hat\theta_B$ is not derived.

{\sc Example 4.} In this example, we evaluate the linear correlation coefficient. We generate $X_1,\cdots,X_N$ and $Y_1,\cdots,Y_N$ from the standard normal distribution and let covariance be $\sigma_{XY}=0.5$. As a result, the population means are $\mu_X=\mu_Y=0$, and the population standard deviations are $\sigma_X=\sigma_Y=1$. The linear correlation coefficient is $\theta=\sigma_{XY}/(\sigma_X\sigma_Y)$. The sample linear correlation based on sequential subsamples can be calculated according to (\ref{eq:hatB}). Similarly, the squared standard error is derived according to (\ref{eq:SEofg}), and the true value of the variance of $\hat\theta_B$ cannot be obtained.

{\sc Example 5.} In the last example, we perform a linear regression. More specifically, we generate the explanatory variables from a multivariate normal distribution with mean 0 and covariance matrix $\Sigma=(\sigma_{j_1j_2})\in\mR^{3\times 3}$. The diagonal elements of $\Sigma$ are 0. For $j_1\not=j_2$, let $\sigma_{j_1j_2}=0.5^{|j_1-j_2|}$. The response is generated according to $Y=\beta_0+\beta_1X_1+\beta_2X_2+\beta_3X_3+\varepsilon$, where the error term $\varepsilon$ is generated from a standard normal distribution. Let $\beta=(\beta_0,\beta_1,\beta_2,\beta_3)^\top=(3,1.5,0,-0.5)^\top$ be the true value of the regression coefficient. Then the ordinary least squares (OLS) estimator can be obtained by $\hat\beta=(X^\top X)^{-1}(X^\top Y)$. The SAS estimator $\hat\theta_B$ is calculated according to (\ref{eq:hatB}), and the squared standard error can be obtained according to (\ref{eq:SEofg}).

\csubsection{Assessment Criteria}

For each simulation setup, various combinations of $(n,B)$ are considered. For each combination, the experiment is randomly replicated $R=$200 times. Let $\bar{\bar X}_B^{(r)}$ or $\hat\theta_B^{(r)}$ be the estimator obtained in the $r$th replication, where $r=1,\cdots,R$. The corresponding estimated squared standard error is calculated according to (\ref{eq:SE}) or (\ref{eq:SEofg}) and we denote it by $\widehat{\mbox{SE}}^{2(r)}$ for the $r$th replication.

We then report the following measurements. First, the mean squared error (MSE) of the estimator is calculated as $R^{-1}\sum_{r=1}^R(\bar{\bar X}_B^{(r)}-\mu)^2$ or $R^{-1}\sum_{r=1}^R(\hat\theta_B^{(r)}-\theta)^2$. Second, the sample variance of $\bar{\bar X}_B$ or $\hat\theta_B$ is calculated across $R$ replications and is denoted by Var. Accordingly, the theoretical variance is denoted by Var$^{*}$, which is $\sigma^2(n^{-1}B^{-1}+N^{-1})$ for the sample mean and $\dot g^2(\mu)\sigma^2(n^{-1}B^{-1}+N^{-1})$ for $\hat\theta_B$. The ratio of the sample variance and the theoretical variance is reported to evaluate the correctness of theorems. Third, the squared standard error (SE$^2$) is calculated as $R^{-1}\sum_{r=1}^R \widehat{\mbox{SE}}^{2(r)}$. We consider the ratio of SE$^2$ and Var$^{*}$ to evaluate the performance of the automatic inference procedure.
It is worth noting that for the last three examples, the theoretical variance of $\hat\theta_B$ is not derived. As a result, Var$^{*}$ is not reported and compared. Instead, we report the ratio of Var and SE$^{2}$.

\csubsection{Simulation Results}

The simulation results for examples 1 and 2 are presented in Table 1. Since Table 1 and Table 2 exhibit similar patterns, we focus on the results in Table 1 for detailed explanations. From Table 1, we obtain the following conclusions. First, the MSE of $\bar{\bar X}_B$ decreases as $n$ or $B$ increases. This means that $\bar{\bar X}_B$ is a consistent estimator of $\mu$. It is worth noting that if $N=10^6$ and $n=10^4$, the increase in $B$ will not lead to a sharp decrease in the MSE. The reason is that in this scenario, the leading order of $\var(\bar{\bar X}_B)$ is $O(N^{-1})$, and the increase in $B$ will not cause this term to be diminished. Second, the ratio of Var and Var$^*$ is approximately 1, which means that the theoretical variance of $\bar{\bar X}_B$ in Theorem 1 is consistent with the sample variance. Third, the ratio of SE$^2$ and Var$^*$ is approximately 1, which means that $\wh{\mbox{SE}}^2(\bar{\bar X}_B)$ is a consistent estimator of $\var(\bar{\bar X}_B)$. Fourth, Table 1 also reports the standard deviation of SE$^2$/Var$^{*}$. The fluctuation of the ratio is mainly determined by $O(B^{-1})$ or $O(n^2N^{-2})$ according to Theorem 3. If $O(B^{-1})$ is the leading order, SD decreases along with the increase of $B$. However, if $O(n^2N^{-2})$ becomes the leading order, an increase in $B$ will not lead to an apparent decrease in SD, e.g., for $N=10^4$ and $n=10^3$.

The simulation results for examples 3-5 are shown in Table 2 and Table 3. In these three examples, the parameter of interest is not just a function of the population mean $\mu$ but a more complex function of the population moments. Since results for these examples exhibit similar patterns, we focus on  the results for example 3 to provide detailed explanations. The following conclusions can be drawn. First, the MSE of the estimator decreases as long as $n$, $B$ or $N$ increases. This means that the SAS based sample CV is a consistent estimator of the population CV. Second, the theoretical variance of the SAS based sample CV is not derived. Hence, Var$^{*}$ cannot be computed and evaluated in this example. Instead, we report the ratio of SE$^2$ and Var. Recall that SE$^2$ is the statistic for automatic inference based on the SAS procedure. Furthermore, Var is the sample variance of $\hat\theta_B$, i.e., the sample CV, derived from $R$ replications. This ratio is approximately 1, which means that the statistic $\wh{\mbox{SE}}^2(\hat\theta_B)$ that we have proposed for automatic inference is reasonable.

\csubsection{Hard Drive Sampling Cost}

In the second part of the experiment, we try to compare the HDSC values obtained for the SAS and RAS methods. Recall that the RAS method is just the so-called $m$ out of $n$ bootstrap but performed on the hard drive. We simulate independent $X_1,\cdots,X_N$ from a standard normal distribution with $N=10^8$. To estimate the population mean $\mu$, the sample mean $\bar{\bar X}_B$ is calculated based on the SAS and RAS subsamples respectively. We repeat the experiment $R=100$ times. Then, the average MSE is reported for both subsampling methods across $R$ replications. Furthermore, HDSC measured in seconds is recorded for one particular combination of $(n,B)$, and the average HDSC is reported for both methods.

The simulation results are shown in Table 4. From Table 4, we draw the following conclusions. First, the MSE values of both methods decrease with increasing $n$ or $B$. If $n$ is large (i.e., $n=10^6$), the MSE values of the SAS method are smaller than those of the RAS method.
Second, the difference between the HDSC values of the SAS and RAS methods is very large. The HSDC value of the RAS method is almost ten times that of the SAS method. For instance, if $n=10^4$ and $B=100$, it takes only 0.4 second for the SAS method to complete the subsampling procedure, while the time needed by the RAS method is almost 6 seconds. To summarize, the SAS and RAS methods have comparable accuracy in terms of MSE. However, the SAS method requires less time if data are massive.

\csection{REAL DATA ANALYSIS}

To demonstrate the practical performance of the proposed SAS method, we perform an empirical study using a large-scale dataset, i.e., the U.S. airline data. The dataset, accessible at {\it http://stat-computing.org/dataexpo/2009}, consists of hundreds of millions of records of commercial flights from October 1987 to April 2008. Each record contains detailed information of a flight. For illustration, we select arrival delay, representing the duration of delay, as the response variable. Since the distribution of arrival delay is heavily tailed at both ends, we filter out the negative values and apply logarithmic transformation to the response. Ultimately, we are left with $5.75\times10^7$ observations. We further choose departure time and day for week as covariates. The departure time is binned into 4 categories, referred to as morning (7am-12pm, used as the base group), afternoon (12pm-6pm), evening (6pm-12am) and midnight (12am-7am), respectively. For categorical variable ``day of week", Monday is treated as the base group.

After shuffling, the dataset takes up to 1.4GB, which is hard to read into memory as a whole. Furthermore, even more memory will be required if one tries to perform statistical analysis on the whole dataset. We then read the data part-by-part and calculate the OLS estimator using the divide-and-conquer approach. More specifically, each block of data contains approximately $10^6$ observations, and the entire dataset is divided into 58 blocks. To total time needed to separately read these 58 blocks of data into memory is approximately 20 seconds. The corresponding OLS estimator is then denoted by $\hat\beta_{\mbox{ols}}$.
We further apply the SAS and RAS methods to the data to derive subsamples. Various combinations of $(n,B)$ combinations are tried, and the SAS and RAS regression coefficient estimators are obtained by averaging the results obtained for $B$ subsamples. Furthermore, the estimated standard error is calculated according to (\ref{eq:hatB}). We also record the HDSC values for both methods.

The detailed results are reported in Table 5. To save space, only the estimates for the intercept and the departure time are reported. From Table 5, we draw the following conclusions. First, the SAS and RAS estimates are very close to the OLS estimates. Second, as $n$ and $B$ increase, the standard errors of both SAS and RAS estimates decrease. Third, the HDSC values of the SAS and RAS methods are largely different. Consider $(n,B)=(10^4,10^3)$ as an example. It takes only approximately 3 seconds for the SAS method to obtain 100 subsamples from the hard drive. However, more than 40 seconds are needed for the RAS method to complete the same procedure. Furthermore, it takes approximately 20 seconds to read all of 58 blocks into memory to obtain the OLS estimation.

\csection{CONCLUSION}

In this paper, we develop a brand-new subsampling method called SAS. This newly proposed method performs subsampling directly from the hard drive, and it can be applied to analyzing massive data under memory constraints. The foundation of the SAS method is sequential subsampling. As a result, it is time saving in terms of HDSC. We further develop the SAS estimators such as the sample mean and the transformation of the sample moments. It is then natural to construct the estimator of the standard error, which can be used for automatic inference. We derive theoretical results for SAS estimators and perform a number of simulation studies. A real example of airline data is also presented.

We note the following possible future research topics. First, the main aim of this paper is to propose a method of sequential subsampling from the hard drive. The SAS estimators are not complex, and are functions of sample moments. More complicated SAS estimators and their properties can be studied in the future. Second, in the real data analysis, we construct a linear regression model, and the OLS estimator is merely a function of sample moments. A generalized linear regression such as the logistic regression can be studied within the framework of the SAS method. Third, it cannot be denied that feature screening and variable selection have been the most popular topics in the past decade. How to conduct the feature screening or variable selection based on the SAS method is another interesting topic for future studies.

\csection{APPENDIX}
\renewcommand{\theequation}{A.\arabic{equation}}
\setcounter{equation}{0}

 We denote by $E^*(\cdot)$, $\var^*(\cdot)$ and $\cov^*(\cdot)$ the conditional expectation, conditional variance, and conditional covariance given $\mB$ throughout the proof.

\scsubsection{Appendix A. Proof of Lemma 1}

We derive the variance of $\bar{\bar{X}}$ in this lemma. Note that $\var(\bar{\bar X})=E(\bar{\bar X}^2)-E^2(\bar{\bar X})$. Since $\bar{\bar X}$ is an unbiased estimator of $\mu$, we only have to derive $E(\bar{\bar X}^2)$.

By the definition of $\bar X_{k}$, it can be easily derived that $E(\bar X_{k}^2)= E^2(\bar X_{k})+\var (\bar X_{k})=\mu^2 + \sigma^2/n$. By the definition of $\bar{\bar{X}}$, we have that
$$ E\big(\bar{\bar{X}}^2\big)=K^{-2}E\Big(\bar X_{1}+\cdots+\bar X_{K}\Big)^2
                    =K^{-2}\bigg\{\sum_{k=1}^{K} E\big(\bar X_{k}^2\big)+\sum_{k_1\not=k_2}E\big(\bar X_{k_1}\bar X_{k_2}\big)\bigg\}.$$
It can be easily derived that $\sum_{k=1}^K E(\bar X_k^2)=K(\mu^2+\sigma^2/n)$. We next calculate the other term $\sum_{k_1\not=k_2}E(\bar X_{k_1}\bar X_{k_2})$. Note that we assume that $n/N\rightarrow 0$. Thus, $K=N-n+1$ should be much larger than the subsample size $n$.

First, if $|k_1-k_2|\geq n$, $E(\bar X_{k_1}\bar X_{k_2})=E(\bar X_{k_1})E(\bar X_{k_2})=\mu^2$. There are $(K-n)(K-n+1)$ pairs of $\bar X_{k_1}$ and $\bar X_{k_2}$ satisfying $|k_1-k_2|\geq n$. As a result, $\sum_{|k_1-k_2|\geq n}E(\bar X_{k_1}\bar X_{k_2})=(K-n)(K-n+1)\mu^2$.
Second, let $|k_1-k_2|=k'$. If $0<k'<n$, $E(\bar X_{k_1}\bar X_{k_2})=\mu^2+n^{-2}(n-k')\sigma^2$, and there are $2(K-k')$ pairs of $\bar X_{k_1}$ and $\bar X_{k_2}$ satisfying
$|k_1-k_2|=k'<n$. As a result, $\sum_{0<|k_1-k_2|< n}E(\bar X_{k_1}\bar X_{k_2})=(2K-n)(n-1)\mu^2+2n^{-2}C\sigma^2$, where $C=n(n-1)(3K-n-1)/6$.
It can then be derived that $E(\bar{\bar{X}}^2)=\mu^2+(nK)^{-2}(nK+2C)\sigma^2$, where the leading term is $\mu^2+\sigma^2/N$ as long as $n\rightarrow\infty$, $N\rightarrow\infty$, and $n/N\rightarrow 0$. We then obtain
\beqrs
\var\big(\bar{\bar X}\big)&=&\frac{(N-n+1)+(n-1)(3N-4n+2)/3}{n(N-n+1)^2}\sigma^2\\
&=&\frac{1}{N}\Bigg\{1+\frac{2n}{3N}+o\Big(\frac{n}{N}\Big)\Bigg\}\sigma^2.
\eeqrs
This completes the proof.

\scsubsection{Appendix B. Proof of Theorem 1}

Since $\var(\bar{\bar{X}}_B)=E\{\var^*(\bar{\bar{X}}_B)\}+\var\{E^*(\bar{\bar{X}}_B)\}$, we next study the conditional expectation and conditional variance of $\bar{\bar{X}}_B$ separately.

First of all, we study the expectation of conditional variance. It can be derived that
\beqr
E\Big\{\var^*\big(\bar{\bar{X}}_B\big)\Big\}&=&B^{-1}E\Big\{\var^*\big(\bar X_{(b)}\big)\Big\} =B^{-1}E\bigg\{E^*\big(\bar X_{(b)}-\bar{\bar{X}}\big)^2\bigg\}\nonumber\\
&=&(BK)^{-1}E\bigg\{\sum_{k=1}^{K}\Big(\bar X_k-\bar{\bar{X}}\Big)^2\bigg\}\nonumber\\
&=&B^{-1}\Big\{E\big(\bar X_k^2\big)-E\big(\bar{\bar X}^2\big)\Big\}.\label{eq:ExpConVar}
\eeqr

Next, we study the variance of conditional expectation. It can be derived that
$$
E^*\Big(\bar{\bar{X}}_B\Big)=B^{-1}\sum_{b=1}^BE^*\Big\{\bar X_{(b)}\Big\}=K^{-1}\sum_{k=1}^K\bar X_k=\bar{\bar{X}}.
$$
Hence,
\beq
\var\Big\{E^*(\bar{\bar X}_B)\Big\}=\var(\bar{\bar X})=E(\bar{\bar X}^2)-E^2(\bar{\bar X}).\label{eq:VarConExp}
\eeq
Combining (\ref{eq:ExpConVar}) and (\ref{eq:VarConExp}) together, we obtain
\beqr
\var(\bar{\bar X}_B)&=&E\Big\{\var^*\big(\bar{\bar{X}}_B\big)\Big\}+\var\Big\{E^*(\bar{\bar X}_B)\Big\}\nonumber\\
&=&B^{-1}\Big\{E\big(\bar X_k^2\big)-E\big(\bar{\bar X}^2\big)\Big\}+E(\bar{\bar X}^2)-E^2(\bar{\bar X})\nonumber\\
&=&(nB)^{-1}\sigma^2+(1-B^{-1})\var(\bar{\bar X})\label{eq:varXB}
\eeqr
From (\ref{eq:barbarx}), we know that
\beqrs
\var(\bar{\bar X}_B)
&=&\frac{\sigma^2}{nB}+(1-B^{-1})\frac{1}{N}\Bigg\{1+O\Big(\frac{n}{N}\Big)\Bigg\}\sigma^2\\
&=&\frac{\sigma^2}{nB}+\frac{1}{N}\Bigg\{1+O\Big(\frac{n}{N}\Big)+O\Big(\frac{1}{B}\Big)\Bigg\}\sigma^2,
\eeqrs
as long as $n\rightarrow\infty$, $N\rightarrow\infty$, $B\rightarrow\infty$, and $n/N\rightarrow 0$.
This completes the proof.

\scsubsection{Appendix C. Proof of Theorem 2}

We study the expectation and bias of $\wh{\mbox{SE}}^2$ in the following two steps. In step 1, the expectation is studied, and in step 2 the bias is derived.

\noindent
{\sc Step 1. Expectation of $\wh{\mbox{SE}}^2$}.

By the definitions of $\bar{\bar{X}}$ and $\bar{\bar{X}}_{B}$, we obtain
\beqr
\wh{\mbox{SE}}^2&=&\frac{c}{B-1} \sum_{b=1}^B \Big(\bar X_{(b)}-\bar{\bar{X}}+\bar{\bar{X}}-\bar{\bar{X}}_{B}\Big)^2\nonumber\\
&=&\frac{c}{B-1}\bigg\{\sum_{b=1}^B\big(\bar X_{(b)}-\bar{\bar{X}}\big)^2-B\big(\bar{\bar{X}}-\bar{\bar{X}}_{B}\big)^2\bigg\}
=\frac{c}{B-1}(A_1-A_2),\label{eq:B12}
\eeqr
where $c=n\{(nB)^{-1}+N^{-1}\}$ can be regarded as a scaler. We next study the expectations of $A_1$ and $A_2$ in (\ref{eq:B12}) separately.
Given $\mB$, all the $\bar X_{(b)}$s can be seen as independent variables. Then, we can derive that
\beqr
E(A_1)&=&E\bigg\{\sum_{b=1}^B\big(\bar X_{(b)}-\bar{\bar{X}}\big)^2\bigg\}=E\bigg[E^*\Big\{\sum_{b=1}^B(\bar X_{(b)}-\bar{\bar{X}})^2\Big\}\bigg]\nonumber\\
&=&B E\bigg\{E^*\Big(\bar X_{(b)}-\bar{\bar{X}}\Big)^2\bigg\}=BK^{-1}E\bigg\{\sum_{k=1}^{K}\Big(\bar X_k-\bar{\bar{X}}\Big)^2\bigg\},\label{eq:B1}
\eeqr
and
\beqr
E(A_2)&=&E\bigg\{B\big(\bar{\bar{X}}-\bar{\bar{X}}_{B}\big)^2\bigg\}=E\bigg[E^*\Big\{B\big(\bar{\bar{X}}_{B}-\bar{\bar{X}}\big)^2\Big\}\bigg]
=E\bigg\{\var^*\Big(\bar X_{(b)}\Big)\bigg\}\nonumber\\
&=& E\bigg\{E^*\big(\bar X_{(b)}-\bar{\bar{X}}\big)^2\bigg\}=K^{-1}E\bigg\{\sum_{k=1}^{K}\Big(\bar X_k-\bar{\bar{X}}\Big)^2\bigg\}.\label{eq:B2}
\eeqr
Using (\ref{eq:B1}) and (\ref{eq:B2}) back to (\ref{eq:B12}), we obtain
\beqr E \big(\wh{\mbox{SE}}^2\big)&=&\frac{c}{B-1}E(A_1-A_2)=\frac{c}{K}E\bigg\{\sum_{k=1}^{K}\Big(\bar X_k-\bar{\bar{X}}\Big)^2\bigg\}
=c\Big\{E\big(\bar X_k^2\big)-E\big(\bar{\bar X}^2\big)\Big\}\nonumber\\
&=&n\Big(\frac{1}{nB}+\frac{1}{N}\Big)\Big\{n^{-1}\sigma^2-\var(\bar{\bar X})\Big\}\label{eq:expSE}\\
&=&\Big(\frac{1}{nB}+\frac{1}{N}\Big)\Bigg\{1-\frac{1}{N-n+1}-\frac{(n-1)(3N-4n+2)/3}{(N-n+1)^2}\Bigg\}\sigma^2\nonumber\\
&=&\Big(\frac{1}{nB}+\frac{1}{N}\Big)\Bigg\{1+O\Big(\frac{n}{N}\Big)\Bigg\}\sigma^2\nonumber.
\eeqr

\noindent
{\sc Step 2. Bias of $\wh{\mbox{SE}}^2$}.

Since $\wh{\mbox{SE}}^2$ is used to estimate the variance of $\bar{\bar X}_B$, we study the bias of this estimator.
Using (\ref{eq:varXB}) and (\ref{eq:expSE}), the bias can be calculated as
\beq
\var\big(\bar{\bar X}_B\big)-E\big(\wh{\mbox{SE}}^2\big)=\Big(1+\frac{n}{N}\Big)\var\big(\bar{\bar X}\big)-\frac{1}{N}\sigma^2
=O\Big(\frac{n}{N^2}\Big)\sigma^2.\label{eq:bias}
\eeq
This completes the proof.

\scsubsection{Appendix D. Proof of Theorem 3}

In this proof, we study the variance of $\wh{\mbox{SE}}^2$. Since $\var(\wh{\mbox{SE}}^2)=E\{\var^*(\wh{\mbox{SE}}^2)\}+\var\{E^*(\wh{\mbox{SE}}^2)\}$, we need to
study the two terms $E\{\var^*(\wh{\mbox{SE}}^2)\}$ and $\var\{E^*(\wh{\mbox{SE}}^2)\}$ separately in step 1 and step 2.

\noindent
{\sc Step 1. Expectation of conditional variance}.

We first study the expectation of conditional variance. First, note that
\beqr
\var^*\Big\{\sum_{b=1}^B\big(\bar X_{(b)}-\bar{\bar X}_B\big)^2\Big\}&=&
\var^*\Big\{\sum_{b=1}^B\big(\bar X_{(b)}-\mu+\mu-\bar{\bar X}_B\big)^2\Big\}\nonumber\\
&=&\var^*\bigg\{\sum_{b=1}^B\big(\bar X_{(b)}-\mu\big)^2-B\big(\bar{\bar X}_B-\mu\big)^2\bigg\}=\var^*\big(B_1-B_2\big)\nonumber\\
&=&\var^*(B_1)+\var^*(B_2)-2\cov^*(B_1,B_2).\label{eq:ConVar}
\eeqr
We next study the terms $B_1=\sum_{b=1}^B(\bar X_{(b)}-\mu)^2$ and $B_2=B(\bar{\bar X}_B-\mu)^2$ separately.

Given $\mB$, all the $\bar X_{(b)}$s are independent. Hence, $\var^*(B_1)=B\var^*(\bar X_{(b)}-\mu)^2$. Then, we obtain
\beqr
E\Big\{\var^*(B_1)\Big\}&=&BE\Big\{\var^*\big(\bar X_{(b)}-\mu\big)^2\Big\}
=BE\bigg[E^*\big(\bar X_{(b)}-\mu\big)^4-\Big\{E^*\big(\bar X_{(b)}-\mu\big)^2\Big\}^2\bigg]\nonumber\\
&=&BE\bigg[K^{-1}\sum_{k=1}^K\big(\bar X_k-\mu\big)^4-\Big\{K^{-1}\sum_{k=1}^K(\bar X_k-\mu)^2\Big\}^2\bigg]\nonumber\\
&=&BE\big(\bar X_k-\mu\big)^4-BK^{-2}E\Big\{\sum_{k=1}^K\big(\bar X_k-\mu\big)^2\Big\}^2=B_{11}-B_{12}.\label{eq:ExpVarA1}
\eeqr
We first study $B_{11}=BE(\bar X_k-\mu)^4$. Note that
\beqrs
E\big(\bar X_k-\mu\big)^4&=&E\Big\{n^{-1}\sum_{i=1}^n(X_i-\mu)\Big\}^4\\
&=&n^{-4}\Big\{n E(X_i-\mu)^4+3n(n-1)E^2(X_i-\mu)^2\Big\}.
\eeqrs
The second equation holds because $E(X_i-\mu)=0$. As a result,
$E(\bar X_k-\mu)^4=3n^{-2}\sigma^4-(3-\gamma)n^{-3}\sigma^4=3n^{-2}\sigma^4\{1+o(1)\}$, and we know that
$B_{11}=3Bn^{-2}\sigma^4\{1+o(1)\}$.

We next study $B_{12}$. Note that
\beqr
E\Big\{\sum_{k=1}^K\big(\bar X_k-\mu\big)^2\Big\}^2&=&E\Big\{\sum_{k=1}^K(\bar X_k-\mu)^4
+\sum_{k_1\not=k_2}(\bar X_{k_1}-\mu)^2(\bar X_{k_2}-\mu)^2\Big\}\nonumber\\
&=&3Kn^{-2}\sigma^4\{1+o(1)\}+K(K-1)n^{-2}\sigma^4\{1+o(1)\}\nonumber\\
&=&K^2n^{-2}\sigma^4+2Kn^{-2}\sigma^4\{1+o(1)\}.\label{eq:B12}
\eeqr
As a result, $B_{12}=Bn^{-2}\sigma^4\{1+o(1)\}$. Substituting $B_{11}$ and $B_{12}$ into (\ref{eq:ExpVarA1}), we obtain
\beq
E\Big\{\var^*(B_1)\Big\}=2Bn^{-2}\sigma^4\{1+o(1)\}.
\eeq

We next study $E\{\var^*(B_2)\}$. Note that
\beqrs
E\Big\{\var^*(B_2)\Big\}&=&B^2E\Big\{\var^*(\bar{\bar X}_B-\mu)^2\Big\}\leq B^2 E\Big\{E^*\big(\bar{\bar X}_B-\mu\big)^4\Big\}\\
&=&B^2E\big(\bar{\bar X}_B-\mu\big)^4=O(n^{-2})=o\Big(E\big\{\var^*(B_1)\big\}\Big).
\eeqrs
The last equation holds if $B\rightarrow\infty$. Similarly, we can prove that $E\{\cov^*(B_1,B_2)\}$ is $o(E\{\var^*(B_1)\})$.
Consequently, we take these results back to (\ref{eq:ConVar}), and since $\wh{\mbox{SE}}^2=c(B-1)^{-1}\sum_{b=1}^B(\bar X_{(b)}-\bar{\bar X}_B)^2$, we can conclude that
\beqr
E\Big\{\var^*\big(\wh{\mbox{SE}}^2\big)\Big\}&=&\frac{c^2}{(B-1)^2}E\Big\{\var^*(B_1)\Big\}\big\{1+o(1)\big\}\nonumber\\
&=&\frac{2\sigma^4}{B}\Big(\frac{1}{nB}+\frac{1}{N}\Big)^2\big\{1+o(1)\big\}.\label{eq:ExpVarSE}
\eeqr

\noindent
{\sc Step 2. variance of conditional expectation}.

From Theorem 1, we know that $E^*(\wh{\mbox{SE}}^2)=cK^{-1}\sum_{k=1}^K(\bar X_k-\bar{\bar X})^2$. As a result,
\beqrs
\var\Big\{E^*(\wh{\mbox{SE}}^2)\Big\}&=&c^2\var\Big\{K^{-1}\sum_{k=1}^K\big(\bar X_k-\bar{\bar X}\big)^2\Big\}\\
&=&c^2\var\bigg\{K^{-1}\sum_{k=1}^K\big(\bar X_k-\mu\big)^2-\big(\bar{\bar X}-\mu\big)^2\bigg\}=c^2\var\big(C_1-C_2\big)\\
&=&c^2\Big\{\var(C_1)+\var(C_2)-2\cov(C_1,C_2)\Big\}.
\eeqrs
We next study the variances of $C_1$ and $C_2$.

First, $E(\bar X_k-\mu)^2=\var(\bar X_k)=\sigma^2/n$. Furthermore, by (\ref{eq:B12}) we know that
\beqrs
\var(C_1)&=&K^{-2}\bigg[E\Big\{\sum_{k=1}^K(\bar X_k-\mu)^2\Big\}^2-E^2\Big\{\sum_{k=1}^K(\bar X_k-\mu)^2\Big\}\bigg]\\
&=&2K^{-1}n^{-2}\sigma^4\{1+o(1)\}.
\eeqrs
Second, $\var(C_2)\leq E(\bar{\bar X}-\mu)^4=O(N^{-2})$. The covariance of $C_1$ and $C_2$ is ignorable compared with the variances of $C_1$
and $C_2$.
As a result, the variance of the conditional expectation is
\beqr
\var\Big\{E^*(\wh{\mbox{SE}}^2)\Big\}&=&n^2\Big(\frac{1}{nB}+\frac{1}{N}\Big)^2\Big\{\var(C_1)+\var(C_2)\Big\}\{1+o(1)\}\nonumber\\
&=&2\sigma^4\Big(\frac{1}{nB}+\frac{1}{N}\Big)^2\Bigg\{O\Big(\frac{1}{N}\Big)+O\Big(\frac{n^2}{N^2}\Big)\Bigg\}.\label{eq:VarExpSE}
\eeqr
By (\ref{eq:ExpVarSE}) and (\ref{eq:VarExpSE}), we know that the variance of $\wh{\mbox{SE}}^2$ should be
$$2\sigma^4\Big(\frac{1}{nB}+\frac{1}{N}\Big)^2
\Bigg\{O\Big(\frac{1}{B}\Big)+O\Big(\frac{1}{N}\Big)+O\Big(\frac{n^2}{N^2}\Big)\Bigg\}.$$
This completes the proof.

\scsubsection{Appendix E. Proof of Theorem 4}

We study the expectation and variance of $\tilde\theta_B$ and $\hat\theta_B$ separately.

\noindent
{\sc Step 1. expectation and variance of $\tilde\theta_B$.}

Using the standard Taylor expansion, we obtain
$\tilde\theta_B=g(\bar{\bar X}_B)=g(\mu)+\dot{g}(\mu)(\bar{\bar X}_B-\mu)+\ddot{g}(\xi_B)(\bar{\bar X}_B-\mu)^2$.
Recall that, from Theorem 1,
\beqr
\var(\bar{\bar X}_B)=\sigma^2\Big(\frac{1}{nB}+\frac{1}{N}\Big)\big\{1+o(1)\big\}.
\eeqr
Since $|\ddot g(\cdot)|\leq C_{\max}$, we then have that
$$E(\tilde\theta_B)=g(\mu)+\sigma^2O\Big(\frac{1}{nB}+\frac{1}{N}\Big).$$
As a result, $\tilde\theta_B$ is a biased estimator of $g(\mu)$, and the bias is $O(n^{-1}B^{-1}+N^{-1})$.
Furthermore, $\var(\bar{\bar X}_B-\mu)^2\leq E(\bar{\bar X}_B-\mu)^4\leq c^*(nB)^{-2}$, where $c^*$ is some constant.
As a result, the variance of $\tilde\theta_B$ can be derived as
\beqrs
\var(\tilde\theta_B)&\leq&\dot g^2(\mu)\var(\bar{\bar X}_B-\mu)+C_{\max}^2\var(\bar{\bar X}_B-\mu)^2+M\\
&=&\dot{g}^2(\mu)\sigma^2\Big(\frac{1}{nB}+\frac{1}{N}\Big)\big\{1+o(1)\big\}.
\eeqrs
The term $M$ is the covariance, which is ignorable compared with the variance terms.

\noindent
{\sc Step 2. Expectation and variance of $\hat\theta_B$.}

From (\ref{eq:TEbarxk}) and the definition of $\hat\theta_B$, we know that
\beqr
\hat\theta_B&=&g(\mu)+\dot g(\mu)B^{-1}\sum_{b=1}^B(\bar X_{(b)}-\mu)+
B^{-1}\sum_{b=1}^B\Big\{\ddot g(\xi_{(b)})(\bar X_{(b)}-\mu)^2\Big\}\nonumber\\
&=&g(\mu)+R_{1B}+R_{2B},\label{eq:R12}
\eeqr
where $\xi_{(b)}$ is between $X_{(b)}$ and $\mu$. We then study $R_{1B}$ and $R_{2B}$ in (\ref{eq:R12}) separately.

First, we define $R_{1B}=\dot g(\mu)B^{-1}\sum_{b=1}^B(\bar X_{(b)}-\mu)=\dot g(\mu)(\bar{\bar X}_B-\mu)$. Then, the expectation of $R_{1B}$ is 0. Furthermore, the variance of $R_{1B}$ is
$$\var(R_{1B})=\dot g^2(\mu)\var(\bar{\bar X}_B)=\dot g^2(\mu)\sigma^2\Big(\frac{1}{nB}+\frac{1}{N}\Big)\big\{1+o(1)\big\}.$$
As a result, $R_{1B}$ is $O_p(\sqrt{(nB)^{-1}+N^{-1}})$.
We next examine $R_{2B}$. Based on condition (C1), we know that
$$E|R_{2B}|\leq C_{\max}B^{-1}\sum_{b=1}^BE\big(\bar X_{(b)}-\mu\big)^2=C_{\max}\sigma^2/n.$$
By condition (C2), $R_{2B}$ is ignorable compared with $R_{1B}$.

The expectation of $\hat\theta_B$ is
\beqrs
E(\hat\theta_B)&=&g(\mu)+B^{-1}\sum_{b=1}^BE\Big\{\ddot g(\xi_{(b)})(\bar X_{(b)}-\mu)^2\Big\}\\
&=&g(\mu)+O\Big(\frac{\sigma^2}{n}\Big).
\eeqrs
We can conclude that $\hat\theta_B$ is also a biased estimator, where the bias is $O(n^{-1})$.
Furthermore, the variance of $R_{2B}$ is $O(n^{-2})$. Thus, the variance of $\hat\theta_B$ is
\beqr
\var(\hat\theta_B)=\var(R_{1B})+O(n^{-2})
=\dot g^2(\mu)\sigma^2\Big(\frac{1}{nB}+\frac{1}{N}\Big)\big\{1+o(1)\big\}.
\eeqr

\scsubsection{Appendix F. Proof of Theorem 5}

According to (\ref{eq:TEbarxk}), we have that
\beqrs
Q&=&\sum_{b=1}^B\Big\{g\big(\bar X_{(b)}\big)-\hat\theta_B\Big\}^2
=\sum_{b=1}^B\Big\{g\big(\bar X_{(b)}\big)-B^{-1}\sum_{m=1}^Bg\big(\bar X_{(m)}\big)\Big\}^2\\
&=&B^{-2}\sum_{b=1}^B\sum_{m=1}^B\Big\{g\big(\bar X_{(b)}\big)-g\big(\bar X_{(m)}\big)\Big\}^2\\
&=&B^{-2}\sum_{b=1}^B\sum_{m=1}^B\Big\{\dot g(\mu)\big(\bar X_{(b)}-\bar X_{(m)}\big)+
\ddot g\big(\xi_{(b)}\big)(\bar X_{(b)}-\mu)^2-\ddot g\big(\xi_{(m)}\big)(\bar X_{(m)}-\mu)^2\Big\}^2\\
&=&B^{-2}\sum_{b=1}^B\sum_{m=1}^B(E_1+E_2)^2.
\eeqrs
Since $E_2$ is ignorable compared with $E_1$, we approximate $Q$ by $\tilde Q$, where
\beqrs
\tilde Q&=&B^{-2}\sum_{b=1}^B\sum_{m=1}^B\Big\{\dot g(\mu)\big(\bar X_{(b)}-\bar X_{(m)}\big)\Big\}^2\\
&=&\dot g^2(\mu)\sum_{b=1}^B\Big(\bar X_{(b)}-\bar{\bar X}_B\Big)^2
\eeqrs
According to (\ref{eq:SE}), we know that $\wh{\mbox{SE}}^2(\hat\theta_B)$ can be approximated well by
$\dot g^2(\mu)\wh{\mbox{SE}}^2(\bar{\bar X}_B)$. Furthermore, using Theorem 2 and Theorem 3, we obtain
\beqrs
E\Big\{\wh{\mbox{SE}}^2(\hat\theta_B)\Big\}&=&\dot g^2(\mu)\sigma^2\Big(\frac{1}{nB}+\frac{1}{N}\Big)\big\{1+o(1)\big\},\\
\var\Big\{\wh{\mbox{SE}}^2(\hat\theta_B)\Big\}&=&2\dot g^4(\mu)\sigma^4\Big(\frac{1}{nB}+\frac{1}{N}\Big)^2
\Bigg\{O\Big(\frac{1}{B}\Big)+O\Big(\frac{1}{N}\Big)+O\Big(\frac{n^2}{N^2}\Big)\Bigg\}.
\eeqrs
This completes the proof.

\newpage
\scsection{REFERENCES}

\begin{description}
\newcommand{\enquote}[1]{``#1''}
\expandafter\ifx\csname natexlab\endcsname\relax\def\natexlab#1{#1}\fi

\bibitem[{Bickel and Freedman, 1981}]{Bickel and Freedman:1981}
Bickel, P.~J. and Freedman, D.~A. (1981),
\enquote{Some asymptotic theory for the bootstrap,}
\textit{The Annals of Statistics}, 9(6), 1196--1217.

\bibitem[{Bickel et~al., 1997}]{Bickel:1997}
Bickel, P.~J., G$\dot{o}$tze, G., and Zwet, W.~R. (1997),
\enquote{Resampling fewer than $n$ observations: gains, losses, and remedies for losses,}
\textit{Statistica Sinica}, 7, 1--31.

\bibitem[{Efron, 1979}]{Efron:1979}
Efron, B. (1979), \enquote{Bootstrap methods: another look at the jackknife,}
\textit{The Annals of Statistics}, 7(1), 1--26.

\bibitem[Fan et~al., 2017]{fan2017distributed}
Fan, J., Wang, D., Wang, K., and Zhu, Z. (2017).
\newblock Distributed estimation of principal eigenspaces.
\newblock {\em arXiv: Computation}.

\bibitem[{Kleiner et~al., 2014}]{Kleiner:2014}
Kleiner, A., Talwalkar, A., Sarkar, P., and Jordan, M.~I. (2014),
\enquote{A scalable bootstrap for massive data,}
\textit{Journal of the Royal Statistical Society, Series B}, 76(4), 795--816.

\bibitem[Lee et~al., 2015]{lee2015communication-efficient}
Lee, J.~D., Sun, Y., Liu, Q., and Taylor, J. (2015).
\newblock Communication-efficient sparse regression: a one-shot approach.
\newblock {\em arXiv: Machine Learning}.

\bibitem[Liu and Ihler, 2014]{liu2014distributed}
Liu, Q. and Ihler, A.~T. (2014).
\newblock Distributed estimation, information loss and exponential families.
\newblock {\em neural information processing systems}, pages 1098--1106.

\bibitem[{Mason and Newton, 1992}]{Mason and Newton:1992}
Mason, D.~M. and Newton, M.~A. (1992),
\enquote{A rank statistics approach to the consistency of a general bootstrap,}
\textit{The Annals of Statistics}, 20(3), 1161--1624.

\bibitem[Zhang et~al., 2012]{zhang2012communication-efficient}
Zhang, Y., Duchi, J.~C., and Wainwright, M.~J. (2012).
\newblock Communication-efficient algorithms for statistical optimization.
\newblock {\em conference on decision and control}, pages 6792--6792.

\end{description}

\begin{table}
\bc\emph{}
\caption{\label{Table1}Simulation Results for Examples 1 and 2. Values of the mean and the standard deviation
are reported for the ratio of SE$^2$ and Var$^*$. The results are averaged over 200 replications.}
\vspace{0.25cm}
\begin{tabular}{c|cc|cccc}
\hline\hline
    &     &     &     &              & \multicolumn{2}{c}{SE$^2$/Var$^*$} \\
$N$ & $n$ & $B$ & MSE($\times 10^{-5}$) & Var/Var$^*$  & Mean & SD \\
\hline
\multicolumn{7}{c}{EXAMPLE 1}\\
    &100&10&105.69&0.96&0.88&0.39\\
    &100&100&15.41&0.77&1.00&0.19\\
$10^4$&100&1000&9.90&0.89&1.00&0.12\\
    &1000&10&20.95&1.05&0.84&0.45\\
    &1000&100&10.45&0.95&0.94&0.35\\
    &1000&1000&9.34&0.91&0.95&0.35\\
\hline
    &100&10&98.73&0.98&0.85&0.37\\
    &100&100&11.54&1.05&0.98&0.14\\
$10^5$&100&1000&1.88&0.95&0.99&0.05\\
    &1000&10&10.09&0.92&0.86&0.39\\
    &1000&100&1.86&0.93&0.96&0.17\\
    &1000&1000&1.10&1.01&0.98&0.13\\
\hline
    &1000&10&10.92&1.09&0.89&0.38\\
    &1000&100&1.04&0.95&1.00&0.15\\
$10^6$&1000&1000&0.18&0.90&1.00&0.06\\
    &10000&10&0.95&0.87&0.91&0.44\\
    &10000&100&0.22&1.07&0.99&0.19\\
    &10000&1000&0.11&0.99&1.00&0.14\\
\hline\hline
\multicolumn{7}{c}{EXAMPLE 2}\\
    &100&10&32.78&0.89&0.95&0.46\\
    &100&100&7.96&1.11&0.99&0.19\\
$10^4$&100&1000&4.84&0.98&0.99&0.12\\
    &1000&10&5.54&0.94&0.81&0.48\\
    &1000&100&3.43&1.07&0.90&0.34\\
    &1000&1000&2.99&1.02&0.91&0.33\\
\hline
    &100&10&27.09&0.88&0.88&0.44\\
    &100&100&4.20&0.96&0.98&0.15\\
$10^5$&100&1000&2.15&0.98&0.99&0.06\\
    &1000&10&3.90&1.22&0.85&0.38\\
    &1000&100&0.63&1.08&0.96&0.18\\
    &1000&1000&0.33&1.02&0.97&0.12\\
\hline
    &1000&10&3.01&1.00&0.93&0.40\\
    &1000&100&0.38&1.08&0.98&0.15\\
$10^6$&1000&1000&0.09&1.06&0.99&0.06\\
    &10000&10&0.33&1.00&0.91&0.40\\
    &10000&100&0.07&1.16&1.00&0.19\\
    &10000&1000&0.04&1.11&1.00&0.13\\
\hline\hline
\end{tabular}
\ec
\end{table}

\begin{table}
\bc\emph{}
\caption{\label{Table1}Simulation Results for Examples 3 and 4. The results are averaged over 200 replications.}
\vspace{0.25cm}
\begin{tabular}{c|cc|cc|cc}
\hline\hline
    &   &   &\multicolumn{2}{c|}{Example 3}&\multicolumn{2}{c}{Example 4}\\
$N$ &$n$&$B$& MSE($\times 10^{-5}$) & SE$^2$/Var & MSE($\times 10^5$) & SE$^2$/Var \\
\hline
    &100&10&179.80&0.90&60.39&0.93\\
    &100&100&36.22&1.07&11.14&1.03\\
$10^4$&100&1000&23.29&0.94&5.92&1.08\\
    &1000&10&27.40&0.89&9.45&0.97\\
    &1000&100&16.60&0.91&5.54&0.99\\
    &1000&1000&16.13&0.87&5.34&0.95\\
\hline
    &100&10&210.44&0.72&59.39&0.93\\
    &100&100&23.16&1.07&6.21&1.01\\
$10^5$&100&1000&9.50&0.96&1.37&1.16\\
    &1000&10&15.07&0.95&5.94&0.89\\
    &1000&100&2.96&1.03&1.08&1.02\\
    &1000&1000&1.63&1.04&0.58&1.06\\
\hline
    &1000&10&14.04&0.94&5.38&0.99\\
    &1000&100&1.54&1.08&0.58&1.04\\
$10^6$&1000&1000&0.28&1.18&0.12&0.99\\
    &10000&10&1.26&1.13&0.66&0.86\\
    &10000&100&0.29&0.99&0.11&0.97\\
    &10000&1000&0.16&1.03&0.06&0.98\\
\hline\hline
\end{tabular}
\ec
\end{table}

\begin{landscape}
\begin{table}
\bc\emph{}
\caption{\label{Table1}Simulation Results for Example 5. Values of MSE is multiplied by $10^5$. The results are averaged over 200 replications.}
\vspace{0.25cm}
\begin{tabular}{c|cc|cc|cc|cc|cc}
\hline\hline
    &   &   &\multicolumn{2}{c}{$\beta_0$}&\multicolumn{2}{c}{$\beta_1$}&\multicolumn{2}{c}{$\beta_2$}&\multicolumn{2}{c}{$\beta_3$}\\
$N$ &$n$&$B$& MSE& SE$^2$/Var & MSE& SE$^2$/Var &MSE & SE$^2$/Var & MSE & SE$^2$/Var\\
\hline
&100&10&107.99&0.93&155.19&0.87&186.61&0.88&146.73&0.86\\
&100&100&18.50&1.07&23.35&1.20&35.45&0.98&31.01&0.89\\
$10^4$&100&1000&11.04&1.00&14.06&1.09&20.83&0.93&18.06&0.86\\
&1000&10&16.59&1.01&23.36&0.98&29.18&1.01&26.94&0.85\\
&1000&100&10.18&1.01&12.41&1.06&18.13&0.98&17.33&0.79\\
&1000&1000&9.40&1.02&12.49&0.96&16.92&0.96&15.59&0.80\\
\hline
&100&10&82.60&1.13&131.62&0.99&154.24&1.04&122.74&1.07\\
&100&100&11.34&0.98&14.23&1.07&17.83&1.06&12.44&1.23\\
$10^5$&100&1000&2.11&0.97&2.96&0.94&3.66&0.95&2.98&0.94\\
&1000&10&12.07&0.77&13.94&0.93&14.68&1.13&15.17&0.89\\
&1000&100&2.16&0.89&2.71&0.97&3.52&0.93&2.93&0.89\\
&1000&1000&1.17&0.93&1.38&1.04&2.09&0.88&1.66&0.87\\
\hline
&1000&10&9.40&1.04&12.53&0.97&15.78&0.98&14.98&0.85\\
&1000&100&1.26&0.85&1.37&1.07&1.66&1.12&1.38&1.05\\
$10^6$&1000&1000&0.20&1.01&0.26&1.03&0.32&1.06&0.33&0.80\\
&10000&10&1.21&0.79&1.40&0.95&2.21&0.73&1.59&0.83\\
&10000&100&0.18&1.09&0.23&1.14&0.34&0.96&0.24&1.09\\
&10000&1000&0.11&1.00&0.13&1.13&0.20&0.89&0.15&0.98\\

\hline\hline
\end{tabular}
\ec
\end{table}
\end{landscape}

\begin{table}
\bc\emph{}
\caption{Values of MSE and HSCD (in seconds) of the SAS and the RAS methods. The experiment is replicated 100 times.}
\vspace{0.25cm}
\begin{tabular}{cc|cc|cc}
\hline\hline
 &  & \multicolumn{2}{c|}{MSE($\times 10^{-7}$)} & \multicolumn{2}{c}{Time cost($\times 10^{-1}$)} \\
n & B & SAS & RAS & SAS & RAS \\
\hline
1000 & 10 & 912.43 & 1032.28 & 0.07 & 0.63  \\
1000 & 50 & 241.87 & 234.79  & 0.35 & 3.13  \\
1000 & 100 & 142.68 & 99.74  & 0.70 & 6.20 \\
\hline
10000 & 10 & 108.27 & 121.08 & 0.47 & 6.01 \\
10000 & 50 & 15.30 & 17.85 & 2.36 & 30.10  \\
10000 & 100 & 12.27 & 10.45 & 4.72 & 59.91  \\
\hline
100000 & 10 & 8.11 & 10.23 & 5.54 & 60.55 \\
100000 & 50 & 1.94 & 2.35  & 27.36 & 303.82 \\
100000 & 100 & 1.04 & 1.24 & 54.65 & 603.59  \\
\hline\hline
\end{tabular}
\ec
\end{table}

\begin{table}
\bc\emph{}
\caption{Regression coefficient estimates and standard errors obtained by the SAS and RAS methods. Each HDSC value is recorded for a combination of $(n,B)$ in seconds. The ordinary least squares estimate is also reported based on the whole sample.}
\vspace{0.25cm}
\begin{tabular}{cc|c|cc|cc|cc}
\hline\hline
$n$ & $B$ & $\hat\beta_{\mbox{ols}}$ & SAS & RAS & SAS & RAS & SAS & RAS \\
\hline
  &   &   & \multicolumn{2}{c|}{$\hat\beta_0$} & \multicolumn{2}{c|}{$\wh{\mbox{SE}}(\hat\beta_0)\times 10^{-2}$} & \multicolumn{2}{c}{HDSC}\\
1000&100&2.04&2.04&2.07&1.75&1.85&0.30&25.36\\
8000&100&2.04&2.04&2.05&0.58&0.64&2.99&38.10\\
10000&100&2.04&2.05&2.04&0.53&0.50&3.46&41.92\\
10000&1000&2.04&2.04&2.04&0.19&0.18&36.80&406.43\\
\hline
  &   &   & \multicolumn{2}{c|}{$\hat\beta_1$} & \multicolumn{2}{c|}{$\wh{\mbox{SE}}(\hat\beta_1)\times 10^{-2}$} & \multicolumn{2}{c}{HDSC}\\
1000&100&0.22&0.23&0.21&1.56&1.74&0.30&25.36\\
8000&100&0.22&0.22&0.22&0.50&0.59&2.99&38.10\\
10000&100&0.22&0.21&0.22&0.50&0.44&3.46&41.92\\
10000&1000&0.22&0.22&0.22&0.16&0.17&36.80&406.43\\
\hline
  &   &   & \multicolumn{2}{c|}{$\hat\beta_2$} & \multicolumn{2}{c|}{$\wh{\mbox{SE}}(\hat\beta_2)\times 10^{-2}$} & \multicolumn{2}{c}{HDSC}\\
1000&100&0.46&0.46&0.45&1.55&1.62&0.30&25.36\\
8000&100&0.46&0.45&0.45&0.52&0.57&2.99&38.10\\
10000&100&0.46&0.45&0.46&0.50&0.48&3.46&41.92\\
10000&1000&0.46&0.46&0.46&0.16&0.17&36.80&406.43\\
\hline
  &   &   & \multicolumn{2}{c|}{$\hat\beta_3$} & \multicolumn{2}{c|}{$\wh{\mbox{SE}}(\hat\beta_3)\times 10^{-2}$} & \multicolumn{2}{c}{HDSC}\\
1000&100&0.59&0.60&0.58&1.65&1.79&0.30&25.36\\
8000&100&0.59&0.60&0.59&0.56&0.62&2.99&38.10\\
10000&100&0.59&0.59&0.59&0.50&0.46&3.46&41.92\\
10000&1000&0.59&0.59&0.59&0.18&0.18&36.80&406.43\\
\hline\hline
\end{tabular}
\ec
\end{table}

\end{document}